\documentclass{article}

\usepackage{arxiv}

\usepackage[utf8]{inputenc} 
\usepackage[T1]{fontenc}    
\usepackage{hyperref}       
\usepackage{url}            
\usepackage{booktabs}       
\usepackage{amsfonts}       
\usepackage{nicefrac}       
\usepackage{microtype}      
\usepackage{lipsum}		

\usepackage{amsmath}
\newtheorem{theorem}{Theorem}[section]
\usepackage{tikz}
\definecolor{processblue}{cmyk}{0.96,0,0,0}
\usepgflibrary{arrows}
\usetikzlibrary{positioning, quotes}
\usepgflibrary{arrows.meta, shapes}
\usetikzlibrary{shapes,snakes,positioning}
\numberwithin{equation}{section}
\usepackage{subcaption}
\title{A Repairable System Supported by Two Spare Units and Serviced by Two Types of Repairers}

\author{
  Vahid Andalib\\
  Department of Mathematical Sciences\\
  Indiana Universiry-Purdue University Indianapolis\\
  Indianapolis, IN\\
  \texttt{vandalib@iupui.edu} \\
   \And
  Jyotirmoy Sarkar\\
  Department of Mathematical Sciences\\
  Indiana Universiry-Purdue University Indianapolis\\
  Indianapolis, IN\\
  \texttt{jsarkar@iupui.edu} \\
}

\begin{document}
\maketitle

\begin{abstract}
We study a one-unit repairable system, supported by two identical spare units on cold standby, and serviced by two types of repairers. The model applies, for instance, to  ANSI\footnote{American National Standard Institute} centrifugal pumps in a chemical plant. The failed unit undergoes repair either by an in-house repairer within a random or deterministic patience time, or else by a visiting expert repairer. The expert repairs one or all failed units before leaving, and does so faster but at a higher cost rate than the regular repairer. Four models arise depending on the number of repairs done by the expert and the nature of the patience time. We compare these models based on the limiting availability $A_\infty$, and the limiting profit per unit time $\omega$, using semi-Markov processes, when all distributions are exponential. As anticipated, to maximize $A_{\infty}$, the expert should repair all failed units. To maximize $\omega$, a suitably chosen deterministic patience time is better than a random patience time. Furthermore, given all cost parameters, we determine the optimum number of repairs the expert should complete, and the optimum patience time given to the regular repairer in order to maximize $\omega$. 
\end{abstract}

\section{Introduction}

Let us begin with a motivating application of our general model.
Pumps are of paramount importance in the chemical industry as they are essential to transfer highly corrosive and abrasive chemicals through pipes. The most widely used pump is the ANSI centrifugal pump. Some unique risks associated with chemical plants are  abrupt production termination, disastrous plant failure and dangerous environmental interference. These risks result in huge, irrecoverable loses. Therefore, it is critically important to minimize the aforementioned risks by developing a redundant system of multiple, repairable ANSI centrifugal pumps to ensure a very high system availability, while maintaining profitability.

We consider a continuously monitored, one-unit repairable system supported by two other identical units, and serviced by two types of repairers in order to  reduce maintenance cost. A regular in-house repairer may have limited maintenance knowledge, but he is paid less per hour and his continual presence eliminates the overhead expense payable to a visiting expert repairer. Generally, the regular repairer can do minor repairs within a given patience time, and is either incapable of performing more complicated repairs, or is unable to do so within the patience time. The visiting expert repairer, on the other hand, can fix any problem with the failed unit, and she performs the repair faster than the regular repairer. However, her hourly charges are comparatively higher, and she must be paid also a trip charge for each visit. 

This is how the system operates: Initially, one unit is put on operation and the other two units are on cold standby. Upon failure of the operating unit, immediately a spare unit is placed on operation, and the failed unit undergoes repair---first by the regular repair person, and if it is not repaired within the patience time $T$, the visiting expert repair person is called in. We allow either a random patience time (RPT) or a deterministic patience time (DPT). We also call in the expert repairer when the system goes down because all three units are down; that is, the regular repairer is busy fixing a previously failed unit, the patience time is not over yet, but the other two units have successively failed.

However, the two repairers cannot work simultaneously since the repair facility can accommodate only one repairer at a time. In particular, while a repairer is working on a failed unit, should another unit fail, it must await repair. Also, we assume that the benefit of any partial repair done by the regular repair person is forfeited when the expert takes over the job. We also assume that when repair is complete by either repairer, the repaired unit becomes as good as new. 

How long will the expert remain at the repair facility? We consider two possibilities before the expert leaves the system: Either she repairs all failed units while she is visiting, which we call the multiple repair by expert (MRE) policy. Or, she fixes only one failed unit during each visit; and she lets the regular repairer attend to the waiting failed unit(s), if any. This second possibility we call the single repair by expert (SRE) policy.  

Depending on the type of patience time---random or deterministic---and the number of repairs done by the expert---single or multiple---four possible models arise: (1) MRE-RPT, (2) SRE-RPT, (3) MRE-DPT, and (4) SRE-DPT. We evaluate the performance of these four models in terms of limiting availability $A_\infty$ and limiting profit per unit time $\omega$. Under the assumption of continuous monitoring and continuous life- and repair times, the limiting availability exists; and it is defined as the long-run proportion of time the system is up \cite{feller1968introduction}. Likewise, the limiting profit per unit time is defined as the long-run difference between the net revenue earned and the repair cost paid to the repair persons, including a trip charge payable to the expert, all expressed per unit time. 

\cite{bieth2010standby} studies Models (1)-(4), when there is only one spare unit. Assuming exponential life- and repair times, they obtain $A_\infty$ and $\omega$ using the technique of semi-Markov processes (SMP). We extend their results to the case of two spare units. Such an extension is desirable if, for example, $A_\infty$ with only one spare unit falls below an acceptable threshold even when the units are state-of-the-art. Assuming that the engineering side has already done its best to manufacture such crucial units, on behalf of the maintenance team we can further improve $A_\infty$ to  exceed the acceptable threshold  by utilizing one more spare unit. 

We demonstrate that the system with two spare units has higher $A_\infty$ and $\omega$ compared to a system with only one spare unit. For any choice of parameter values, 
we determine a range of values of $T$ for which Model (3) performs the best in terms of both $A_\infty$ and $\omega$. Thus, if we choose $T$ in this range, then the DPT policy, which is logistically preferable to implement, yields higher $A_\infty$ and $\omega$ than the RPT policy. Furthermore, we obtain a threshold value for the cost per unit time payable to the expert repairer such that so long as the expert charges less than this threshold value the MRE policy yields higher profit than the SRE policy, and vice versa.


The rest of the paper is organized as follow: In Section 2, we give a literature review. In Section 3, we formulate the stochastic behavior of the repairable system as an SMP; and we describe the analytic techniques for deriving the limiting availability and the limiting profit per unit time. In Section 4, we provide detailed analytic derivations for all four repair models. Section 5 compares the four models against those when there is only one spare unit. Finally, Section 6 concludes the paper with a summary and several directions for future research.   

\section{Literature Review}
In this section, we review some latest developments in modeling repairable systems to address various reliability characteristics.

\cite{sarkar2006limiting} considers a one-unit repairable system, supported by $r$ identical repair facilities and $s$ cold standby spare units, $r\leq s+1$, which fails when all units are down and are undergoing or awaiting repair. They obtain limiting average availability under a perfect repair policy when lifetime is arbitrary and repair time is exponential. \cite{sarkar2010availability} studies a similar model, but they obtain the instantaneous availability function under both life- and repair times exponentially distributed.

\cite{wang2007reliability} deals with reliability and sensitivity analysis of a repairable system with several operating- and warm standby units, and several unreliable service stations. Failure times and service times are exponentially distributed, and the service station is subject to breakdowns according to a Poisson process. They determine the mean time to failure (MTTF) and system reliability; and study how these characteristics change with the model parameters.

\cite{zhang2007deteriorating} studies a cold standby repairable system consisting of two dissimilar components---with Component 1 having priority in use---and one repairman. Component 2 is as good as new after repair, while Component 1 follows a geometric process repair. Assuming exponential life- and repair times, they derive some important reliability indices such as the system availability, reliability, mean time to first failure (MTTFF), rate of occurrence of failure and the probability the repairman remains idle. For Component 1, they determine an optimal replacement policy which minimizes the long-run average cost per unit time. 

\cite{yu2007optimal} designs a maintainable cold standby system which minimizes the system cost rate subject to availability constraint. \cite{el2010stochastic}  investigates the cost-benefit analysis of a two-unit cold standby system with two-stage repair with waiting time in between. They use regenerative point processes to obtain time dependent availability, steady state availability, reliability, MTTF and profit function.

\cite{cui2017new} proposes two interval availability indexes for Markov repairable systems which measure the probability that the system is working during a given time window containing either a specified point or an interval. \cite{yi2018stochastic} studies a discrete-time semi-Markovian repairable system where the state space of the process includes three subsets---working, changeable and failed. They apply Z-transform to derive reliability, point availability and interval availability. They also discuss for their system the two new reliability measures introduced in \cite{cui2017new}.

\cite{cha2019stochastic} describes repairable systems in which defects are detected before failure, triggering repair. The system is either perfectly repaired within a time period, and the process renews; or it is not repaired within the time period, causing fatal failure. The authors derive the survival function of these systems assuming exponential time to defect, deterministic time period and arbitrary repair time; though they illustrate the results only under exponential repair time. They also obtain asymptotic survival probability under the assumption of fast repair when distributions are arbitrary.        

Repairable systems with two types of repairers have not been studied extensively. \cite{kumar1996comparative} studies Model (2) with only one spare unit. They allow an expert to take over the repair only after the patience time of the regular repairer is exhausted without completing the repair, even if the system fails during this time.  \cite{sridharan1998stochastic} calls in the expert as soon as the patience time is over or the system fails. Although they claim to allow arbitrary life-, repair- and patience time distributions, their results are correct only under exponential life- and exponential repair times, as pointed out in  \cite{bieth2010standby}. \cite{sridharan2000probabilistic} allows a random pre-inspection time for the regular repairer to determine whether he is able to repair a failed unit or not. If he is capable of repairing, he starts the repair; otherwise, the expert is called immediately.  \cite{bieth2010standby} studies Models (1)-(4), when there is only one spare unit. They obtain limiting availability and limiting profit per unit time using the SMP technique under exponential life- and repair times. They also extend the technique to allow arbitrary life- and repair times. 

\cite{parashar2007reliability} studies a one-unit system backed by a hot standby spare unit in a master-slave relationship. Initially, the master unit is operating and the slave unit is on hot standby. There are three types of failures: minor, major-repairable and major-irreparable (which requires replacement). The regular repairer repairs only minor failures. They claim to derive the system MTTF, steady-state availability and limiting profit per unit time assuming repair- and replacement times are arbitrary but lifetime is exponential; however, no analytic solutions are given. In fact, their theoretical results are valid only under exponential life-, repair- and replacement times.

The papers discussed above utilize the Laplace transform technique to obtain various system reliability indices including, but not limited to, availability, busy periods for the two repairers and profit. Since Laplace transforms are often challenging to invert, the technique is not practically implementable. In this paper we derive the limiting results in a straight-forward and simpler manner using semi-Markov processes (SMP).

\section{System Description and Mathematical Framework}
For four models discussed in Section 1, we study the system limiting availability and limiting profit per unit time under the following assumptions:
\begin{enumerate}
	\item A one-unit system has three identical units. At the very beginning, one unit is put on operation, and the other two spare units remain on cold standby.
	\item There is only one repair facility attended by either the regular or the expert repairer.
	\item Failure of the operating unit is immediately detected; the failed unit is sent for repair, and if a standby unit is available, it is put on operation immediately.
	\item The regular repair person has to finish repair within a maximum
	allowable patience time $T$ which may be random (RPT) or deterministic (DPT).
	\item The system fails when all three units are down.
	\item When either the patience time for the regular repair person is over or the system fails, whichever happens first, the expert is called; and she arrives immediately.
	\item When the expert repairer takes over the job, the benefits of partial repair done by the regular repairer is forfeited.
	\item Life-, repair- and patience times are exponentially distributed with arbitrary parameters, and are independent of one another.
	\item We consider two options for the expert repairer: She may leave the repair facility after repairing all failed units, which is called the MRE model. Or,  she may leave the facility after repairing only one failed unit and letting the regular repairer attend to the other failed unit(s), if any. This alternative model is called the SRE model.
	\item We assume a perfect repair policy under which a repaired unit becomes as good as new.
\end{enumerate}
At any time, a unit exhibits one of five possible features: $s$ (on standby), $p$ (operating), $r$ (undergoing repair by regular repairer), $e$ (undergoing repair by expert repairer) or $w$ (awaiting repair). Since the units are identical, it suffices to record how many units are exhibiting each feature. Accordingly, the system is in one of six possible states: $1=(p,s,s)$, $2=(r,p,s)$, $3=(e,p,s)$, $4=(r,w,p)$, $5=(e,w,p)$, $6=(e,w,w)$. The system is down in State 6, and is up in all other states.

\begin{figure}[h]
	\centering
	\tikzset{
		node distance = 15mm and 13mm,
		box/.style = {rectangle, draw=gray!25, thin,
			top color=white, bottom color = gray!50,
			text height=1.5ex, text depth=0.25ex,
			minimum width=6em},
		ellip/.style = {ellipse, draw=gray!25, thin,
			top color=white, bottom color = gray!50},
		sx/.style = {xshift=#1mm},
		every edge quotes/.append style = {font=\footnotesize,
			inner sep=1pt, auto=right},
		every edge/.append style = {-Stealth}
	}
	\begin{tikzpicture}
	\node (n6) [ellip]  {$6=(e,w,w)$};
	\node (n4) [box, above  left=of n6.north]   {$4=(r,w,p)$};
	\node (n5) [box, above right=of n6.north]   {$5=(e,w,p)$};
	\node (n2) [box, above=of n4]               {$2=(r,p,s)$};
	\node (n3) [box, above=of n5]               {$3=(e,p,s)$};
	\node (n1) [box, above=of n2 -| n6]         {$1=(p,s,s)$};
	\draw       ([sx=-6] n1.south)  edge ["$X$"] ([sx=-2] n2.north)
	([sx= 2] n2.north)  edge ["$Y$"] ([sx=-2] n1.south)
	([sx=-2] n2.south)  edge ["$X$"] ([sx=-2] n4.north)
	([sx= 2] n4.north)  edge ["$Y$"] ([sx= 2] n2.south)
	([sx=-2] n4.south)  edge ["$X'$"] (n6.150)
	(n6.120)            edge ["$Z$", dashed] ([sx=3] n4.south)
	([sx=-2] n3.north)  edge ["$Z$"] ([sx= 2] n1.south)
	(n3.south)          edge ["$X$"] (n5.north)
	(n5.south)          edge ["$X$"] (n6.45)
	([sx=-4] n5.north)  edge ["$Z$", dashed] ([sx=6] n2.south)
	(n2)    edge ["$T$"  '] (n3)
	(n4)    edge ["$T'$" '] (n5);
	\end{tikzpicture}
	\hfil
	\begin{tikzpicture}
	\node (n6) [ellip]  {$6=(e,w,w)$};
	\node (n4) [box, above  left=of n6.north]   {$4=(r,w,p)$};
	\node (n5) [box, above right=of n6.north]   {$5=(e,w,p)$};
	\node (n2) [box, above=of n4]               {$2=(r,p,s)$};
	\node (n3) [box, above=of n5]               {$3=(e,p,s)$};
	\node (n1) [box, above=of n2 -| n6]         {$1=(p,s,s)$};
	\draw       ([sx=-6] n1.south)  edge ["$X$"] ([sx=-2] n2.north)
	([sx= 2] n2.north)  edge ["$Y$"] ([sx=-2] n1.south)
	([sx=-2] n2.south)  edge ["$X$"] ([sx=-2] n4.north)
	([sx= 2] n4.north)  edge ["$Y$"] ([sx= 2] n2.south)
	([sx=-2] n4.south)  edge ["$X'$"] (n6.150)
	([sx=6] n6.50)            edge ["$Z$", dashed] ([sx=3.5] n5.south)
	([sx=-2] n3.north)  edge ["$Z$"] ([sx= 2] n1.south)
	(n3.south)          edge ["$X$"] (n5.north)
	([sx=-1]n5.south)          edge ["$X$"] (n6.40)
	([sx=4] n5.north)  edge ["$Z$", dashed] ([sx=4] n3.south)
	(n2)    edge ["$T$"  '] (n3)
	(n4)    edge ["$T'$" '] (n5);
	\end{tikzpicture}
	\caption{Transition diagrams for SRE (left) and MRE (right) models}
	\label{fig:diagrams}
\end{figure}
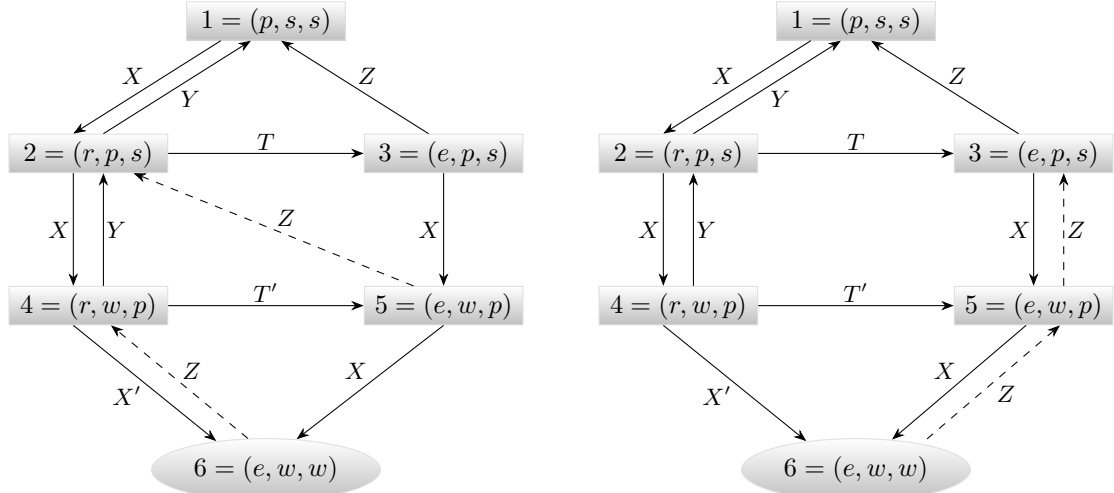

Figure \ref{fig:diagrams}  shows the transitions under SRE and MRE models, along with random variables that determine the sojourn times and transition probabilities.

Let us first explain the random variables. Let $X$, $Y$ and $Z$ denote the lifetime of the unit, the repair time by the regular repairer and the repair time by the expert respectively. Some additional random variables shown in the diagram have the following interpretations: The variable $X'$ is another lifetime which has the same distribution as $X$, but is independent of $X$. The variable $T'$ is the remaining patience time. It  reduces to $T'=T-X$ under the DPT policy; but under the RPT policy, in view of the memoryless property of exponential distribution,  $T'$ has the same distribution as $T$, but it is independent of $T$.

Next, let us explain the sojourn times in each state and the transitions out of them. The system starts in State 1 at time $t=0$; it stays there for a random duration $X$; and then it moves to State 2. The sojourn time in State 2 is $min(X,Y,T)$; and the system returns to State 1 if $Y$ is the smallest, to State 3 if $T$ is the smallest, or to State 4 if $X$ is the smallest. The sojourn time in State 3 is $min(X,Z)$; and the system moves to State 1 if $Z<X$, or to State 5 otherwise. The sojourn time in State 4 is $min(X',Y,T')$; and the system moves to State 2 if $Y$ is the smallest, to State 5 if $T'$ is the smallest, or to State 6 if $X'$ is the smallest. The sojourn time in State 5 is $min(X,Z)$. The system moves to State 6 if $X<Z$; otherwise, it moves to State 3 (under MRE policy) or to State 2 (under SRE policy). Finally, as soon as the expert repairs the failed unit in State 6, the system moves to State 5 (under MRE policy) or to State 4 (under SRE policy). The dashed arrows emphasize the transitions exclusive to each model, while the solid arrows are common to both models. The transition probabilities out of each state are determined based on whichever associated random variable attains the minimum.

Let $\theta_k$ be the proportion of time the system spends in State $k\ (k=1,\ldots, 6)$. Since the system is down in State 6, the limiting availability of the system is,
\begin{equation}\label{eq1}
	A_\infty=1-\theta_6=\theta_1 + \theta_2+\theta_3+\theta_4+\theta_5.
\end{equation}
Having obtained $A_\infty$, we can now derive  $\omega$, the limiting profit per unit time. We need the following parameters: The proportion of busy time for the regular repairer is $\Theta_r=\theta_2+\theta_4$, and that for the expert is $\Theta_e=\theta_3+\theta_5+\theta_6$. Let $R_p, C_p,C_r,C_e$ denote respectively the net revenue, the operation cost, the payment to the regular repairer and the payment to the expert---all expressed per unit time. Also, let $C_l$ denote the trip charge paid to the expert per trip (not per unit time). Then the limiting profit per unit time is given by
\begin{equation}\label{eq2}
	\omega=A_\infty(R_p-C_p)-[\Theta_r C_r+\Theta_e C_e+C_l/\tau],   
\end{equation}
where $\tau$ is the expected length of a cycle, which is defined as the duration from the epoch the system enters State 2, until it returns to State 2 after visiting one of States 3, 5 and 6 at least once. Thus, within each cycle, the expert comes and returns exactly once, and she is paid the trip charge $C_l$ exactly once. By Wald's First Identity \cite{feller1968introduction}, the expected number of visits by the expert per unit time is the reciprocal of $\tau$. Therefore, $C_l/\tau$ is the trip charge paid to the expert per unit time. 
 
\section{Limiting Availability and Limiting Profit Analysis}
In this section, we derive the analytic expressions for the limiting availability $A_\infty$ and the limiting profit per unit time $\omega$ for all four models: (1) MRE-RPT, (2) SRE-RPT, (3) MRE-DPT, and (4) SRE-DPT. In view of Assumption 8, let us denote the patience time, the lifetime, the repair times by the regular repairer and the expert respectively as
\begin{center}
	$T\sim exp(\alpha),\ \ \ X\sim exp(\lambda),\ \ \ Y\sim exp(\beta),\ \ \ Z\sim exp(\gamma)$.
\end{center}
Here, the parameter of an exponential distribution denotes the rate; and its reciprocal denotes the mean. By the memoryless property of an exponential random variable,  the future trajectory of the stochastic process depends only on the present state, while the history of the process can be disregarded. Hence, the process, describing each repair model is a semi-Markov processes (SMP); that is, the system changes states in accordance with a Markov chain, but takes a random amount of time between changes.  See \cite{ross} for more details on SMP. More specifically, in our models, the embedded discrete time stochastic process (DTSP) is a Markov chain with a finite state space $\{1,2,3,4,5,6\}$ and a transition probability matrix $P=((P_{ij}));\ i,j=1,\ldots,6$. The exact expressions for $P_{ij}$ varies across the four models, and will be presented in the respective subsections. 

The stationary distribution of a Markov chain gives the limiting probability $\pi_j$
of transitions entering (also departing) State $j$. It is unique, and is obtained by solving the following system of equations (for more details see \cite{ross}, pp. 175-177),

\begin{equation}\label{eq4}
	\pi_j=\sum_i \pi_iP_{ij}, \ \ \sum_j \pi_j=1. 
\end{equation}Moreover, the expected sojourn times in different states are
\begin{equation}\label{eq5}
	\begin{aligned}
		\mu_1&=E[X]=\frac{1}{\lambda}\\
		\mu_2&=E[min(X, Y, T)]=
		\left\{ \begin{array}{ll}
			{\frac{1}{\lambda+\alpha+\beta}} \ \ \ \ \ \ \ \ \ \ {\rm{ RPT\ Policy }}\\ \\
			{\frac{1-e^{-(\lambda+\beta)T}}{\lambda+\beta}}\ \ \ {\rm{ DPT\ Policy }}
		\end{array}\right. \\
		\mu_3&=E[min(X, Z)]=\frac{1}{\lambda+\gamma}\\
		\mu_4&=E[min(X', Y, T')]=\left\{ \begin{array}{ll}
			{\frac{1}{\lambda+\alpha+\beta}} \hspace{2.96cm} {\rm{ RPT\ Policy }}\\ \\
			{\frac{1}{\lambda+\beta}-\frac{\lambda (e^{\beta T}-1) e^{-(\lambda+\beta)T}}{\beta(\lambda+\beta)}} \ \ \ \ \ {\rm{ DPT\ Policy }}
		\end{array}\right. \\
		\mu_5&=E[min(X, Z)]=\frac{1}{\lambda+\gamma}\\
		\mu_6&=E[Z]=\frac{1}{\gamma}.
	\end{aligned}
\end{equation}
The following theorem gives the proportions of time the SMP spends in the different states.
\begin{theorem}
	For an SMP, if the embedded DTSP is irreducible with stationary probabilities $\pi$, and if the times between successive visits to any State $k$ has a non-lattice distribution with a finite mean, and $\mu_k$ is the expected sojourn time in State $k$ before transition, then the limiting probability that the process will be found in State $k$ exists, is independent of the initial state, and is given by
	\begin{equation}\label{eq3}
		\theta_k=\frac{\pi_k\mu_k}{\sum_{j=1}^{6}\pi_j \mu_j}.
	\end{equation}
\end{theorem}
In the following subsections, for each of the four models, starting from the transition matrix $P$, we derive $\theta_k$ ($k=1,\ldots 6$) using (\ref{eq3}), (\ref{eq4}) and (\ref{eq5}).  Then we obtain $A_\infty$ using (\ref{eq1}). Next, we obtain the analytic expression of $\tau$  in each model by solving a suitable system of recursive relations. Subsequently, we obtain $\omega$  using (\ref{eq2}).

\subsection{Model 1: MRE-RPT}
For the MRE-RPT repair model, the embedded DTMC has transition matrix
\begin{equation}\label{model1matrix}
	P=
	\begin{bmatrix}
		0 & 1 & 0 & 0 & 0 & 0\\
		\frac{\beta}{\lambda+\alpha+\beta} & 0 & \frac{\alpha}{\lambda+\alpha+\beta} & \frac{\lambda}{\lambda+\alpha+\beta} & 0 & 0\\
		\frac{\gamma}{\lambda+\gamma} & 0 & 0 & 0 & \frac{\lambda}{\lambda+\gamma} & 0\\
		0 & \frac{\beta}{\lambda+\alpha+\beta} & 0 & 0 & \frac{\alpha}{\lambda+\alpha+\beta} & \frac{\lambda}{\lambda+\alpha+\beta}\\
		0 & 0 & \frac{\gamma}{\lambda+\gamma} & 0 & 0 & \frac{\lambda}{\lambda+\gamma}\\
		0 & 0 & 0 & 0 & 1 & 0
	\end{bmatrix}.
\end{equation}
Solving the system of equations (\ref{eq4}), we obtain the stationary distribution as
\begin{multline}\label{eq10}
	\pi  \propto \Big(1-\frac{\lambda\beta}{(\lambda+\alpha+\beta)^2},1,\frac{(\lambda+\gamma)((\lambda+\alpha)^2+\alpha\beta)}{\gamma(\lambda+\alpha+\beta)^2},\frac{\lambda}{\lambda+\alpha+\beta},\\
	\frac{(\lambda+\gamma)(\lambda^2(\gamma+\lambda)+\lambda\alpha(\alpha+2\lambda)+\lambda\alpha(\beta+\gamma)}{\gamma^2(\lambda+\alpha+\beta)^2},\\
	\frac{\lambda^4+\lambda\alpha(\lambda\alpha+2\lambda^2)+\lambda^2\alpha(\beta+\gamma)+\lambda^2\gamma(\gamma+\lambda)}{\gamma^2(\lambda+\alpha+\beta)^2}\Big).
\end{multline}
Substituting the mean sojourn times (\ref{eq5}) and the stationary distribution (\ref{eq10}) into (\ref{eq3}), we can obtain expressions for $\theta_k$'s. Thereafter, from (\ref{eq1}), we get
\begin{equation*}
	A_{\infty}=1-\theta_6
\end{equation*}
where
\begin{equation}
	\theta_6 \propto \mu_6\pi_6 = \frac{\lambda^4+\lambda\alpha(\lambda\alpha+2\lambda^2)+\lambda^2\alpha(\beta+\gamma)+\lambda^2\gamma(\gamma+\lambda)}{\gamma^3(\lambda+\alpha+\beta)^2}. 
\end{equation}

Next, the expected length of a cycle satisfies the recursive relation
\begin{equation}\label{tau1}
	\tau=\mu_2+P_{21}(\mu_1+\tau)+P_{23}\sigma_{32}^M+P_{24}\sigma_{42}^M
\end{equation}
where $\sigma_{32}^M$ denotes the expected time for the system to go from State 3 to State 2 (via State 1 or State 5) under the MRE policy. The other parameters $\sigma_{42}^M$ and $\sigma_{52}^M$ (to be introduced shortly) denote similar quantities. These parameters satisfy
\begin{equation}\label{eq12}
	\begin{aligned}
		\sigma_{32}^M&=\mu_3+P_{31}\mu_1+P_{35}\sigma_{52}^M\\
		\sigma_{52}^M&=\mu_5+P_{53}\sigma_{32}^M+P_{46}(\mu_6+\sigma_{52}^M).
	\end{aligned}
\end{equation}
Solving the system of equations  (\ref{eq12}), we obtain
\begin{equation}\label{eq14}
	\sigma_{52}^M=\frac{\mu_5+P_{53}\mu_3+P_{53}P_{31}\mu_1+P_{56}\mu_6}{1-P_{53}P_{35}-P_{56}}.
\end{equation}Thereafter, we also obtain an explicit expression for $\sigma_{32}^M$ from (\ref{eq12}). Finally. we have one more relationship
\begin{equation}\label{sigma42}
	\sigma_{42}^M =\mu_4+P_{45}\sigma_{52}^M+P_{46}(\mu_6+\sigma_{52}^M)+P_{42}\tau.
\end{equation}Substituting the expressions for $\sigma_{32}^M$ and $\sigma_{42}^M$ into  (\ref{tau1}) and solving, we obtain
\begin{equation}\label{eq20}
	\tau=\frac{\mu_2+P_{21}\mu_1+1-P_{23}\sigma_{32}^M+1-P_{24}(\mu_4+P_{45}\sigma_{52}^M+P_{46}(\mu_6+\sigma_{52}^M))}{1-P_{21}-P_{24}P_{42}}.
\end{equation}

Using expression (\ref{eq20}) for $\tau$, we obtain $\omega$ from (\ref{eq2}).

\subsection{Model 2: SRE-RPT}
For the SRE-RPT repair model, the embedded DTMC has transition matrix
\begin{equation}\label{model2matrix}
	P=
	\begin{bmatrix}
		0 & 1 & 0 & 0 & 0 & 0\\
		\frac{\beta}{\lambda+\alpha+\beta} & 0 & \frac{\alpha}{\lambda+\alpha+\beta} & \frac{\lambda}{\lambda+\alpha+\beta} & 0 & 0\\
		\frac{\gamma}{\lambda+\gamma} & 0 & 0 & 0 & \frac{\lambda}{\lambda+\gamma} & 0\\
		0 & \frac{\beta}{\lambda+\alpha+\beta} & 0 & 0 & \frac{\alpha}{\lambda+\alpha+\beta} & \frac{\lambda}{\lambda+\alpha+\beta}\\
		0 & \frac{\gamma}{\lambda+\gamma} & 0 & 0 & 0 & \frac{\lambda}{\lambda+\gamma}\\
		0 & 0 & 0 & 1 & 0 & 0
	\end{bmatrix}.
\end{equation}
Solving the system of equations (\ref{eq4}), we obtain the stationary distribution as
\begin{equation}\label{eq17} 
	\pi \propto \left(\frac{\beta(\lambda+\gamma)+\gamma\alpha}{(\lambda+\gamma)(\lambda+\alpha+\beta)},1,\frac{\alpha}{\lambda+\alpha+\beta},\xi_1,\xi_2,\frac{\lambda\xi_1}{\lambda+\alpha+\beta}+\frac{\lambda\xi_2}{\lambda+\gamma}\right)
\end{equation}
where
\begin{align*}
	\xi_1=\frac{\lambda(\lambda+\gamma)^2+\alpha\lambda^2}{(\lambda+\gamma)[(\alpha+\beta)\lambda+\beta\gamma]}~~{\rm and}~~
	\xi_2=\frac{\alpha\lambda+\alpha(\lambda+\gamma)\xi_1}{(\lambda+\gamma)(\lambda+\alpha+\beta)}.
\end{align*}
Substituting the mean sojourn times (\ref{eq5}) and the stationary distribution (\ref{eq17}) into (\ref{eq3}), we can obtain expressions for $\theta_k$'s. Therefore, from (\ref{eq1}) we get
\begin{equation*}
	A_{\infty}=1-\theta_6
\end{equation*}
where
\begin{equation}
	\theta_6 \propto \mu_6\pi_6 = \frac{\lambda}{\gamma}\left\{\frac{\xi_1}{\lambda+\alpha+\beta}+\frac{\xi_2}{\lambda+\gamma}\right\}.
\end{equation}
To obtain  $\omega$ we need to find the expected cycle time $\tau$. Let $\sigma_{32}^S$ denote the expected time for the system to go from State 3 to State 2 (via State 1 or State 5) under the SRE policy. Let $\sigma_{42}^S$ and $\sigma_{52}^S$ denote similar quantities. They satisfy the recursive relations
\begin{equation}\label{eq8}
	\begin{aligned}
		\tau&=\mu_2+P_{21}(\mu_1+\tau)+P_{23}\sigma_{32}^S+P_{24}\sigma_{42}^S \\
		\sigma_{32}^S&=\mu_3+P_{31}\mu_1+P_{35}\sigma_{52}^S\\	
		\sigma_{42}^S&=\mu_4+P_{45}\sigma_{52}^S+P_{46}(\mu_6+\sigma_{42}^S)+P_{42}\tau\\
		\sigma_{52}^S&=\mu_5+P_{56}(\mu_6+\sigma_{42}^S).
	\end{aligned}
\end{equation}
Substituting the fourth equation into the third in the system of equations (\ref{eq8}), we obtain

\begin{equation}\label{eq6}
	\sigma_{42}^S=\frac{\mu_4+P_{45}\mu_5+P_{45}P_{56}\mu_6+P_{46}\mu_6+P_{42}\tau}{1-P_{45}P_{56}-P_{46}}.
\end{equation}
Substituting  (\ref{eq6}) into the fourth equation in  (\ref{eq8}) we obtain $\sigma_{52}^S$, and then from the second equation in   (\ref{eq8}) we obtain $\sigma_{32}^S$. Having obtained all the $\sigma^S$'s, from the first equation in (\ref{eq8}), we get 
\begin{equation}\label{eq21}
	\tau=\frac{\mu_2+P_{21}\mu_1+P_{23}[\mu_3+P_{31}\mu_1+P_{35}(\mu_5+P_{56}\mu_6+P_{56}\xi_3)]+P_{24}\xi_3}{1-P_{21}-\frac{P_{23}P_{35}P_{56}P_{42}+P_{24}P_{42}}{1-P_{45}P_{56}-P_{46}}},
\end{equation}
where
\begin{equation*}
	\xi_3=\frac{\mu_4+P_{45}\mu_5+P_{45}P_{56}\mu_6+P_{46}\mu_6}{1-P_{45}P_{56}-P_{46}}.
\end{equation*}

Using expression (\ref{eq21}) for $\tau$, we obtain $\omega$ from (\ref{eq2}).
\subsection{Model 3: MRE-DPT}
For the MRE-DPT repair model, the embedded DTMC has transition matrix
\begin{equation}\label{eq15}
	P=
	\begin{bmatrix}
		0 & 1 & 0 & 0 & 0 & 0\\
		\frac{\beta(1-e^{-(\lambda+\beta)T})}{\lambda+\beta} & 0 & e^{-(\lambda+\beta)T} & \frac{\lambda(1-e^{-(\lambda+\beta)T}}{\lambda+\beta} & 0 & 0\\
		\frac{\gamma}{\lambda+\gamma} & 0 & 0 & 0 & \frac{\lambda}{\lambda+\gamma} & 0\\
		0 & P_{42} & 0 & 0 & P_{45} & P_{46}\\
		0 & 0 & \frac{\gamma}{\lambda+\gamma} & 0 & 0 & \frac{\lambda}{\lambda+\gamma}\\
		0 & 0 & 0 & 0 & 1 & 0
	\end{bmatrix}.
\end{equation}
We left unspecified the transition probabilities out of State~4. Let us explain how to obtain them. Write $T'=T-X$ as the remaining patience time when the system enters State~4 from State~2 because the operating unit fails at time $X<T$. Also, write $X'$ as the lifetime of the newly installed unit, and $Y'$ as the remaining repair time while the regular repairer continues to repair the same failed unit. Then $\min\{X', Y'\}$ follows an exponential distribution with parameter $\lambda+\beta$. Hence,
\begin{align*}
	P_{45}&=P\{T'<\min\{X',Y'\}|X<T\} \\
	&= \int_0^T e^{-(\lambda+\beta)(T-x)}\lambda e^{-\lambda x}dx 
	=\frac{\lambda e^{-\lambda T}(1-e^{-\beta T})}{\beta}.
\end{align*}
Thereafter, we have $P_{42}=(1-P_{45})\beta/(\lambda+\beta)$ and  $P_{46}=(1-P_{45})\lambda/(\lambda+\beta)$.
Solving the system of equations (\ref{eq4}), we obtain the stationary distribution as 
\begin{equation}\label{eq18}
	\pi \propto \left(\frac{\lambda+\beta-\lambda P_{42}(1-e^{-(\lambda+\beta)T})}{\lambda+\beta},1,\xi_4, \xi_5,\frac{P_{46}\lambda(1-e^{-(\lambda+\beta)T})}{\lambda+\beta}+\frac{\lambda\xi_5}{\lambda+\gamma}\right)
\end{equation} 
where
\begin{align*}
	\xi_4&=\frac{\lambda+\gamma}{\gamma(\lambda+\beta)}\Big[\lambda+\beta-(1-e^{-(\lambda+\beta)T})(\beta+\lambda P_{42})\Big] \\
	\xi_5&=\frac{\lambda}{\gamma}\xi_4+\frac{\lambda(\lambda+\gamma)}{\gamma(\lambda+\beta)}(1-e^{-(\lambda+\beta)T})(P_{45}+P_{46}).
\end{align*}
Substituting the mean sojourn times (\ref{eq5}) and stationary distribution (\ref{eq18}) into (\ref{eq3}), we can obtain expressions for $\theta_k$'s. Thereafter, from (\ref{eq1}) we get
\begin{equation*}
	A_{\infty}=1-\theta_6
\end{equation*}
where
\begin{equation}
	\theta_6 \propto \mu_6\pi_6 = \frac{P_{46}\lambda(1-e^{-(\lambda+\beta)T})}{\gamma(\lambda+\beta)}+\frac{\lambda\xi}{\gamma(\lambda+\gamma)}.
\end{equation}
Moreover, $\tau$ satisfies relations similar to (\ref{tau1}) and (\ref{sigma42}) derived in  the MRE-RPT model; but now it uses transition matrix (\ref{eq15}) instead of (\ref{model1matrix}). The corresponding solution for $\tau$ is also similar in form to (\ref{eq20}); but it uses $P_{ij}$'s from (\ref{eq15}).

Using this new expression for $\tau$, we obtain $\omega$ from (\ref{eq2}).
\subsection{Model 4: SRE-DPT}
Finally, for the SRE-DPT repair model, the embedded DTMC has transition matrix
\begin{equation}\label{eq16}
	P=
	\begin{bmatrix}
		0 & 1 & 0 & 0 & 0 & 0\\
		\frac{\beta(1-e^{-(\lambda+\beta)T})}{\lambda+\beta} & 0 & e^{-(\lambda+\beta)T} & \frac{\lambda(1-e^{-(\lambda+\beta)T}}{\lambda+\beta} & 0 & 0\\
		\frac{\gamma}{\lambda+\gamma} & 0 & 0 & 0 & \frac{\lambda}{\lambda+\gamma} & 0\\
		0 & P_{42} & 0 & 0 & P_{45} & P_{46}\\
		0 & \frac{\gamma}{\lambda+\gamma} & 0 & 0 & 0 & \frac{\lambda}{\lambda+\gamma}\\
		0 & 0 & 0 & 1 & 0& 0
	\end{bmatrix}
\end{equation}
where $P_{42}$, $P_{45}$, $P_{46}$ are exactly the same as those in the MRE-DPT model. Solving the system of equations (\ref{eq4}), we obtain the stationary distribution as
\begin{equation}\label{eq19}
	\pi \propto \left(\xi_6,1,e^{-(\lambda+\beta)T},\xi_7,\frac{\lambda e^{-(\lambda+\beta)T}}{\lambda+\gamma}+P_{45}\xi_7,
	\Big(P_{46}+\frac{\lambda P_{45}}{\lambda+\gamma}\Big)\xi_7+\frac{\lambda^2e^{-(\lambda+\beta)T}}{(\lambda+\gamma)^2}\right)
\end{equation}
where
\begin{align*}
	\xi_6&=\frac{\beta}{\lambda+\beta}\big(1-e^{-(\lambda+\beta)T}\big)+\frac{\gamma }{\lambda+\gamma}e^{-(\lambda+\beta)T}
	\\
	\xi_7&=\frac{\frac{\lambda}{\lambda+\beta}\big(1-e^{-(\lambda+\beta)T}\big)+\big(\frac{\lambda}{\lambda+\gamma}\big)^2e^{-(\lambda+\beta)T}}{\frac{\beta}{\lambda+\beta}(1-P_{45})-\frac{\gamma}{\lambda+\gamma}P_{45}}.
\end{align*}
Substituting the mean sojourn times (\ref{eq5}) and stationary distribution (\ref{eq19}) into (\ref{eq3}), we obtain expressions for $\theta_k$'s. Thereafter, from (\ref{eq1}) we get
\begin{equation*}
	A_{\infty}=1-\theta_6
\end{equation*}
where
\begin{equation}
	\theta_6 \propto \mu_6\pi_6 = \frac{\xi_6}{\gamma}\left(\frac{\lambda}{\lambda+\beta} \big(1-P_{45}\big) +\frac{\lambda}{\lambda+\gamma}P_{45}\right)+\frac{\lambda^2e^{-(\lambda+\beta)T}}{\gamma(\lambda+\gamma)^2}.
\end{equation}
Furthermore, $\tau$ satisfies relations similar to  (\ref{eq8}) derived in the SRE-RPT model; but now it uses transition matrix (\ref{eq16}) instead of (\ref{model2matrix}). The corresponding solution for $\tau$ is similar in form to (\ref{eq21}), but it uses $P_{ij}$'s from (\ref{eq16}).

Using this new expression for $\tau$, we obtain $\omega$ from (\ref{eq2}).

\section{Comparison of Models}
In this section, for some choices of values of the parameters, we compare the four repair models discussed in Section 3 in terms of the limiting availability $A_\infty$ and the limiting profit per unit time $\omega$. For a given choice of parameter values, we determine the best model under which both criteria are maximized. We also demonstrate that a system with two spare units has a higher $A_\infty$ and a higher $\omega$  than a system with only one spare unit. 

\begin{figure}
	\centering
	\includegraphics[scale=0.35]{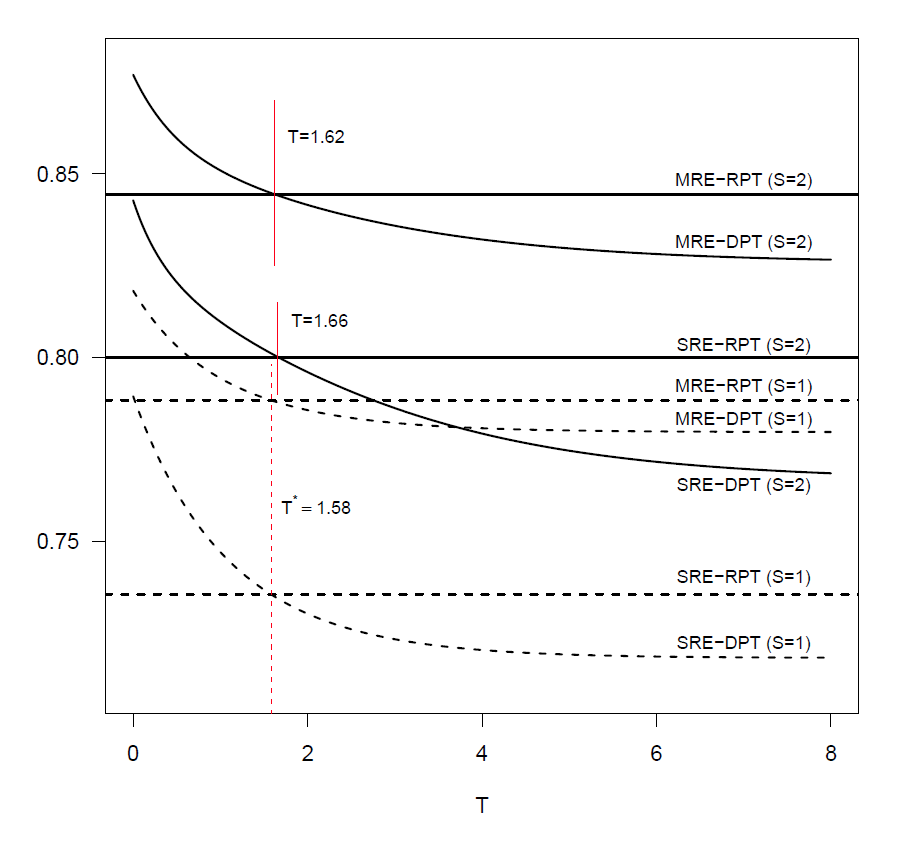}
	\caption{Limiting availability under all repair models for systems with one spare unit (dashed) and two spare units (solid).}
	\label{fig:availability}
\end{figure}

Figure \ref{fig:availability} depicts $A_\infty$ as a function of the patience time $T$ under all four repair models for the systems with either one spare unit ($S=1$) or two spare units ($S=2$) for parameter values: $\lambda=0.5$, $\alpha=0.3$ (RPT), $\beta=0.35$ and $\gamma=0.75$. 

We observe the following results:
\begin{enumerate}
	\item The limiting availability $A_\infty$ is strictly higher under MRE policy than under SRE policy for systems with either one or two spare units, irrespective of the type of patience time adopted. 
	\item As $T \rightarrow \infty$, $A_\infty$ decreases under DPT policy for both MRE and SRE models. Likewise, as $\alpha \rightarrow 0$,  $A_\infty$ decreases under RPT policy. 
	\item Adding one more spare unit to a system supported by only one spare unit, increases $A_\infty$ under both RPT and DPT policies. For example, in the RPT case, $A_\infty$ is below 80\% when $S=1$; but it is more than 80\% when $S=2$. 
	\item Suppose that $S=1$. The choice of $T$, which causes $A_\infty$ to be the same (or equivalently, that causes $\mu_2$ to be the same) under both RPT and DPT policies, is given by (see \cite{bieth2010standby} for further details)
	\begin{equation*}
	\begin{aligned}
	T^*&=\frac{ln(1+\frac{\lambda+\beta}{\alpha})}{\lambda+\beta}
	\end{aligned}
	\end{equation*} 
	For our choice of parameter values, the corresponding $T^*=1.58$. The common value of $A_\infty$ for both RPT and DPT policies is 0.74 for SRE models and 0.79 for MRE models.  
	\item Suppose that $S=2$. The explicit expressions for the choice of $T$ which causes $A_\infty$ to be the same under both RPT and DPT policies is too cumbersome to display. For our choice of parameter values,  we find $T=1.62$ and $A_\infty=0.84$ for MRE models; and $T=1.66$ and $A_\infty=0.80$ for SRE models. 
\end{enumerate}

Thus, under the limiting availability criterion alone, for the system supported by two spare units, the MRE-DPT model is the best, so long as the patience time is not too long, namely $T\leq 1.62$. This is in agreement with the result of \cite{bieth2010standby} for the one spare unit system.
\begin{figure}
	\centering
	\includegraphics[scale=0.35]{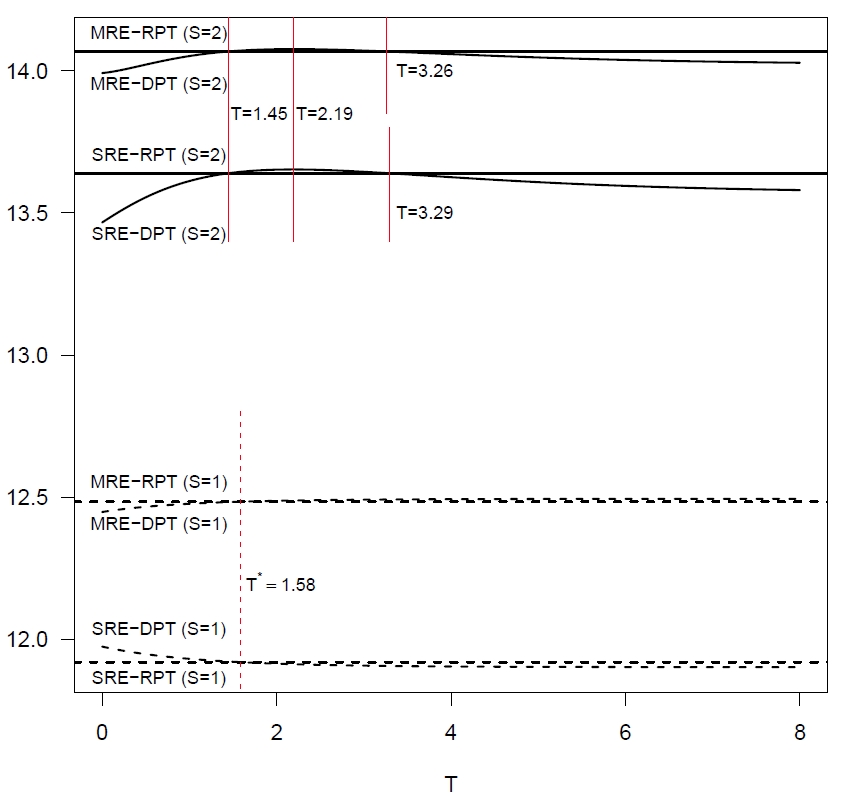}
	\caption{Limiting profit per unit time under all repair models for the systems with one spare unit (dashed) and two spare units (solid).}
	\label{fig:profit}
\end{figure}

Next, we compare the models in terms of the limiting profit per unit time criterion. We assume that the expert repairer completes repair quicker than the regular repairer, but she charges a higher rate; that is, $\beta<\gamma$ and $C_r<C_e$. Figure \ref{fig:profit} depicts $\omega$ as a function of patience time $T$ under all four repair models for systems supported by one spare unit ($S=1$) or two spare units ($S=2$), given the same parameter values as above, and additionally: $R=20$, $C_r=1$, $C_e=5$ and $C_l=3$.

For our choice of parameter values,we observe the following results: 
\begin{enumerate} 
	\item  The limiting profit per unit time $\omega$ is strictly larger under MRE policy than under SRE policy for both cases $S=1$ and $S=2$.
	\item  As $T \rightarrow \infty$, under $S=1$, $\omega$ increases (decreases) slightly under MRE (SRE). However, under $S=2$, $\omega$ first increases in $T$, and then decreases marginally for both MRE and SRE models under DPT policy. In general, as $\alpha \rightarrow 0$ under RPT policy, $\omega$ increases. 
	\item Adding one more spare unit to the system backed by only one spare unit, increases $\omega$ in all four models considerably. 
	\item Under $S=1$, $\omega$ is the same (11.92) for SRE-RPT and SRE-DPT models at  $T^*=1.58$, and it is the same (12.48) for MRE-RPT and MRE-DPT models. However, under $S=2$, $\omega$ is the same (14.07) for MRE-RPT and MRE-DPT models at two time points--- $T=1.45$ and $T=3.26$; and it is the same (13.64) for SRE-RPT and SRE-DPT models at two time points--- $T=1.45$ and $T=3.29$. Hence, for any choice of $T$ in the range $[1.45,3.26]$, $\omega$ is higher under DPT model than under the RPT model. This suggests MRE-DPT as the best model under the limiting profit per unit time criterion (for our choice of parameter values).  
	\item Furthermore, under $S=2$, $\omega$ is maximized at $T=2.19$ under both SRE and MRE models, reaching 13.65 and 14.08 respectively. 
\end{enumerate}

Considering both the limiting availability and the limiting profit per unit time criteria simultaneously, we conclude that for any  choice of $T$ in the range $[1.45,1.62]$, the highest values for both $A_\infty$ and $\omega$ are attained by the MRE-DPT model. The knowledge of this optimum range of values for the patience time $T$ is crucial for maintenance engineers to accomplish management objectives.

\begin{figure}
	\centering
	\begin{subfigure}{7cm}
		\centering\includegraphics[scale=0.3]{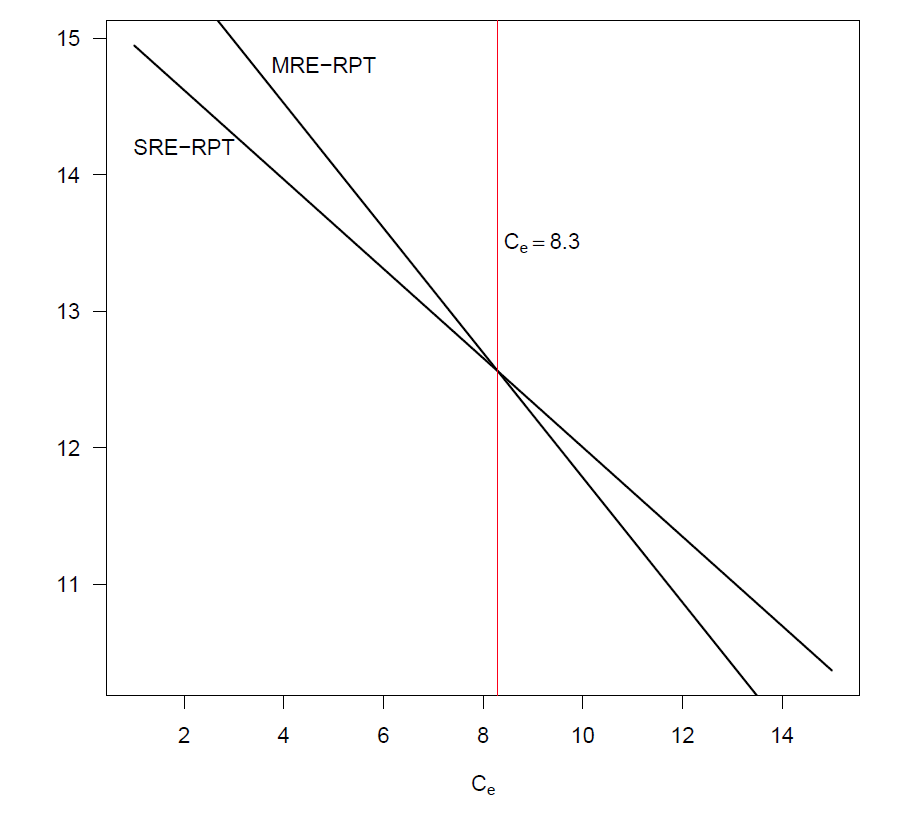}
		\caption{}
	\end{subfigure}
	\begin{subfigure}{7.5cm}
		\centering\includegraphics[scale=0.3]{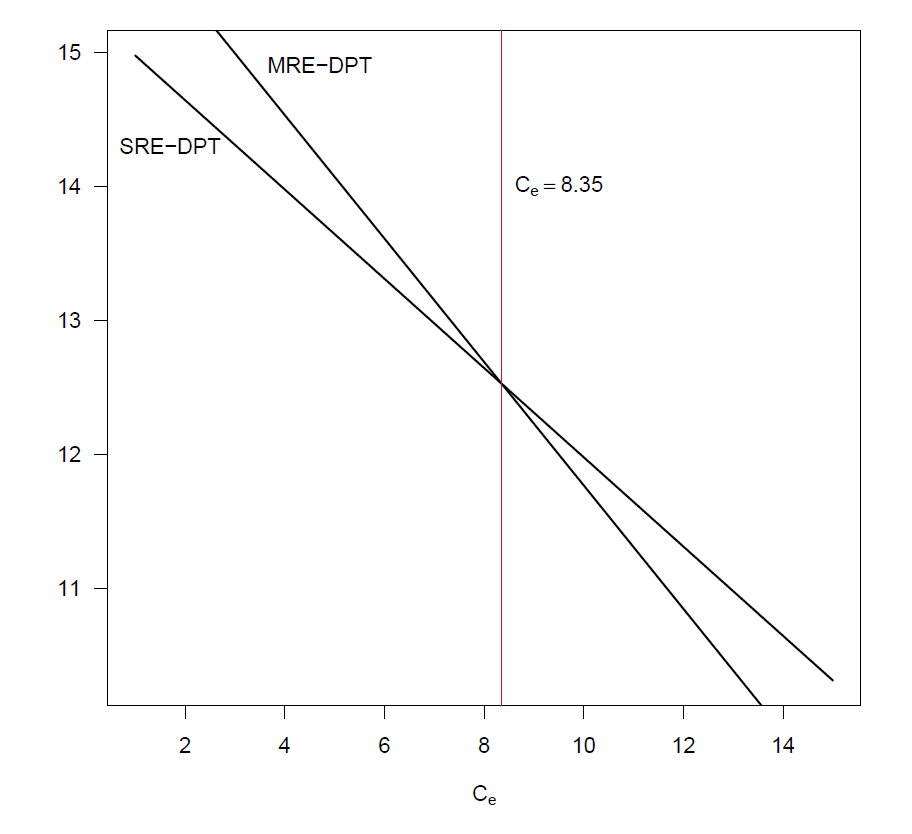}
		\caption{}
	\end{subfigure}
	\caption{Limiting profit per unit time as a function of $C_e$: (a) under RPT policy, (b) under DPT policy with $T=1.5$.}
	\label{fig:omega}
\end{figure}

Although, for our choice of parameter values, it was seen that $\omega$ is larger under MRE policy than under SRE policy, if the expert charges too much, then MRE model may not dominate SRE model in terms of $\omega$. Figure \ref{fig:omega} depicts $\omega$ for MRE and SRE models as the cost per unit time paid to the expert repairer $C_e$ varies with $R=20, C_r=1, C_l=3$. If the expert charges at a rate less than a threshold, then MRE model yields a higher limiting profit per unit time than SRE model under RPT policy; and th opposite holds if the expert charges above the threshold. See panel (a). A similar result  holds under the DPT policy. See panel (b). 

\section{Concluding Remarks}
In this paper, we extend \cite{bieth2010standby} by adding another spare unit to a cold standby repairable system consisting of two identical units and serviced by two types of repair persons. In a situation where component lifetime is short and repair time is long, multiple spare units are necessary to improve the reliability characteristics of the system. In this extended set up, we study the limiting availability and the limiting profit per unit time when lifetime and repair times are exponentially distributed. Four possible models arise depending on the number of failed units the expert repairer is allowed to repair during each visit and on the type of patience time for the regular repairer. We derive the limiting availability and limiting profit per unit time for each of the four possible models using SMP, which is much simpler than the Laplace transform technique widely used in the literature. We show that the system supported by two spare units results in higher $A_\infty$ and higher $\omega$ compared to the system having only one spare unit. 

As in \cite{bieth2010standby}, in our extended set up also a logistically easier to implement DPT model yields higher $A_\infty$ and higher $\omega$ than an RPT model, provided $T$ is chosen appropriately. 
Since the expert repairs faster than the regular repairer, MRE yields a higher $A_{\infty}$ than SRE. However, in order to maximize $\omega$, the maintenance administrator may adopt either MRE or SRE policy depending on the relative costs payable to the expert (compared to the regular repairer).
Thus, given all cost parameters, the maintenance engineer can determine whether MRE or SRE is the preferred policy in terms of $\omega$, and obtain an optimum value of the patience time $T$ that  maximizes $\omega$. 

In our motivating example of ANSI centrifugal pumps in a chemical plant, the maintenance engineer can make decisions on how many repairs the expert should do during each visit and how much patience time should be given the regular repairer to ensure higher limiting availability and limiting profit per unit time. Such informed decisions will minimize any potential economic, health and environmental risks associated with the chemical plant.

We identify several directions of future research:\\
(i) For the purpose of building the repairable models, we have assumed life- and repair times to be exponential. Relaxing these assumptions, though desirable, may prove to be challenging since the stochastic process will no longer be an SMP. \\ \\ 
(ii)  We assumed that there is only one repair facility that allows only one repairer to work at a time. It will be advantageous to employ two repair facilities so that both repairers can work at the same time. Under this assumption, the transition diagram becomes more complicated involving more states. In addition, the Markovian property fails under the DPT policy, since the transition out of some states may depend not only on the current state but also on the history of the process.
\\ \\
(iii) We assumed that the units are identical. It is desirable to study a more realistic system involving non-identical units with different life- and repair rates. In particular, we must determine which unit should be put on operation and which on repair whenever there are multiple such units. \\ \\

\bibliographystyle{unsrt}  
\bibliography{references}  

\begin{thebibliography}{10}

\bibitem{feller1968introduction}
William Feller.
\newblock {\em An introduction to probability theory and its applications},
  volume~1.
\newblock Wiley, New York, 1968.

\bibitem{bieth2010standby}
Bruno Bieth, Liang Hong, and Jyotirmoy Sarkar.
\newblock A standby system with two types of repair persons.
\newblock {\em Applied Stochastic Models in Business and Industry},
  26(5):577--594, 2010.
\newblock \url{https://doi.org/10.1002/asmb.801}.

\bibitem{sarkar2006limiting}
Jyotirmoy Sarkar and Fang Li.
\newblock Limiting average availability of a system supported by several spares
  and several repair facilities.
\newblock {\em Statistics \& probability letters}, 76(18):1965--1974, 2006.
\newblock \url{https://doi.org/10.1016/j.spl.2006.04.046}.

\bibitem{sarkar2010availability}
Jyotirmoy Sarkar and Atanu Biswas.
\newblock Availability of a one-unit system supported by several spares and
  repair facilities.
\newblock {\em Journal of the Korean Statistical Society}, 39(2):165--176,
  2010.
\newblock \url{https://doi.org/10.1016/j.jkss.2009.05.001}.

\bibitem{wang2007reliability}
Kuo-Hsiung Wang, Jyh-Bin Ke, and Wen-Chiung Lee.
\newblock Reliability and sensitivity analysis of a repairable system with warm
  standbys and r unreliable service stations.
\newblock {\em The International Journal of Advanced Manufacturing Technology},
  31(11-12):1223--1232, 2007.
\newblock \url{ https://doi.org/10.1007/s00170-005-0298-0}.

\bibitem{zhang2007deteriorating}
Yuan~Lin Zhang and Guan~Jun Wang.
\newblock A deteriorating cold standby repairable system with priority in use.
\newblock {\em European Journal of Operational Research}, 183(1):278--295,
  2007.
\newblock \url{https://doi.org/10.1016/j.ejor.2006.09.075}.

\bibitem{yu2007optimal}
Haiyang Yu, Farouk Yalaoui, {\.E}ric Ch{\^a}telet, and Chengbin Chu.
\newblock Optimal design of a maintainable cold-standby system.
\newblock {\em Reliability Engineering \& System Safety}, 92(1):85--91, 2007.
\newblock \url{https://doi.org/10.1016/j.ress.2005.11.001}.

\bibitem{el2010stochastic}
Khaled~M El-Said and Mohamed~S El-Sherbeny.
\newblock Stochastic analysis of a two-unit cold standby system with two-stage
  repair and waiting time.
\newblock {\em Sankhya B}, 72(1):1--10, 2010.
\newblock \url{https://doi.org/10.1007/s13571-010-0001-9}.

\bibitem{cui2017new}
Lirong Cui, Jianhui Chen, and Bei Wu.
\newblock New interval availability indexes for markov repairable systems.
\newblock {\em Reliability Engineering \& System Safety}, 168:12--17, 2017.
\newblock \url{https://doi.org/10.1016/j.ress.2017.03.016}.

\bibitem{yi2018stochastic}
He~Yi, Lirong Cui, Jingyuan Shen, and Yan Li.
\newblock Stochastic properties and reliability measures of discrete-time
  semi-markovian systems.
\newblock {\em Reliability Engineering \& System Safety}, 176:162--173, 2018.
\newblock \url{https://doi.org/10.1016/j.ress.2018.04.014}.

\bibitem{cha2019stochastic}
Ji~Hwan Cha and Maxim Finkelstein.
\newblock Stochastic modeling for systems with delayed failures.
\newblock {\em Reliability Engineering \& System Safety}, 2019.
\newblock \url{https://doi.org/10.1016/j.ress.2019.03.017}.

\bibitem{kumar1996comparative}
A~Kumar, SK~Gupta, and G~Taneja.
\newblock Comparative study of the profit of a two server system including
  patience time and instruction time.
\newblock {\em Microelectronics and reliability}, 36:1595--1601, 1996.
\newblock \url{https://doi.org/10.1016/0026-2714(95)00075-5}.

\bibitem{sridharan1998stochastic}
V~Sridharan and P~Mohanavadivu.
\newblock Stochastic behaviour of a two-unit standby system with two types of
  repairmen and patience time.
\newblock {\em Mathematical and computer modelling}, 28(9):63--71, 1998.
\newblock \url{https://doi.org/10.1016/S0895-7177(98)00145-9}.

\bibitem{sridharan2000probabilistic}
V~Sridharan.
\newblock Probabilistic measures of redundant system with two types of
  repairmen, sensing device and analytical approach to find the optimium
  interchanging time.
\newblock {\em International Journal of Quality \& Reliability Management},
  17(9):984--1002, 2000.
\newblock \url{https://doi.org/10.1108/02656710010353894}.

\bibitem{parashar2007reliability}
Bhupender Parashar and Gulshan Taneja.
\newblock Reliability and profit evaluation of a plc hot standby system based
  on a master-slave concept and two types of repair facilities.
\newblock {\em IEEE Transactions on reliability}, 56(3):534--539, 2007.
\newblock \url{https://doi.org/10.1109/TR.2007.903151}.

\bibitem{ross}
Sheldon Ross.
\newblock {\em Stochastic Processes}.
\newblock Wiley, New York, 1966.

\end{thebibliography}






\end{document}